\documentclass[pra,nofootinbib,twocolumn,amsmath,amssymb,superscriptaddress]{revtex4-2}
\usepackage{graphicx,amsmath,relsize,epstopdf,color,mathtools,bm,newtxtext,newtxmath,braket,rotating}
\usepackage[hyphenbreaks]{breakurl}
\usepackage[colorlinks=true,linkcolor=blue,citecolor=blue,urlcolor =blue]{hyperref}
\usepackage[normalem]{ulem}

\usepackage{booktabs}

\usepackage[table,xcdraw]{xcolor}
\newcommand{\eq}[1]{\begin{equation}\begin{aligned}#1\end{aligned}\end{equation}}
\newcommand{\expct}[1]{\left\langle#1\right\rangle}

\newcommand{\eu}{\text{e}}
\newcommand{\ha}{\hat{a}}
\newcommand{\had}{\hat{a}^\dagger\vphantom{a}}

\usepackage{soul,xcolor}

\usepackage{dsfont}

\begin{document}
	\setstcolor{red}

	\title{Optimal transmission estimation with dark counts}
	
	\author{Aaron Z. Goldberg}
    \affiliation{National Research Council of Canada, 100 Sussex Drive, Ottawa, Ontario K1N 5A2, Canada}
    \affiliation{Department of Physics, University of Ottawa, 25 Templeton Street, Ottawa, Ontario, K1N 6N5 Canada}
    \author{Khabat Heshami}
    \affiliation{National Research Council of Canada, 100 Sussex Drive, Ottawa, Ontario K1N 5A2, Canada}
    \affiliation{Department of Physics, University of Ottawa, 25 Templeton Street, Ottawa, Ontario, K1N 6N5 Canada}
    \affiliation{Institute for Quantum Science and Technology, Department of Physics and Astronomy, University of Calgary, Alberta T2N 1N4, Canada}

	\begin{abstract}
	    Transmission measurements are essential from fiber optics to spectroscopy. Quantum theory dictates that the ultimate precision in estimating transmission or loss is achieved using probe states with definite photon number and photon-number-resolving detectors (PNRDs). Can the quantum advantage relative to classical probe light still be maintained when the detectors fire due to dark counts and other spurious events? We demonstrate that the answer to this question is affirmative and show in detail how the quantum advantage depends on dark counts and increases with Fock-state-probe strength. These results are especially pertinent as the present capabilities of PNRDs are being dramatically improved.
	    
	\end{abstract}
	
	
	\maketitle
\section{Introduction}
Quantum metrology promises improvements in estimation precision relative to classical probes and measurement devices using a comparable amount of resources \cite{Caves1981,Dowling1998,Giovannettietal2004,Mitchelletal2004,Berryetal2009,LIGO2011,Humphreysetal2013,Tayloretal2013,Tsangetal2016,Liuetal2020,LupuGladsteinetal2022}. 
These advantages are present in transmission estimation and mathematically equivalent problems \cite{JakemanRarity1986,Heidmannetal1987,Hayatetal1999,Abouraddyetal2001,Abouraddyetal2002entangledellipsometry,Toussaintetal2004,Grahametal2006,MonrasParis2007,Brambillaetal2008,Adessoetal2009,MonrasIlluminati2011,Alipouretal2014,Crowleyetal2014,Medaetal2017,Nair2018,Rudnickietal2020,Ioannouetal2021,WangAgarwal2021} and have been experimentally demonstrated on numerous occasions \cite{Tapsteretal1991,SoutoRibeiroetal1997,YabushitaKobayashi2004,DAuriaetal2006,Bridaetal2010,Moreauetal2017,SabinesChesterkingetal2017,Samantarayetal2017,Whittakeretal2017,Loseroetal2018,Shietal2020,Yoonetal2020,Atkinsonetal2021,Woodworthetal2022}, with applications
including ellipsometry \cite{AzzamBashara1977,Fujiwara2007,Tuchin2016}, spectroscopy \cite{Cheongetal1990,Savageetal1996,Hollas2004,Coneetal2015}, and the characterization of quantum devices \cite{Spagnoloetal2014,Wangetal2017,Arrazolaetal2021,GoldbergHeshami2022arxiv}.

The method guaranteed by quantum metrology to be optimal for measuring transmission requires detectors capable of resolving the finest possible energy differences: they must discriminate between different numbers of photons arriving at the detectors \cite{MonrasParis2007,Adessoetal2009}. Such photon-number-resolving detectors (PNRDs) have been maturing over the last few years \cite{Litaetal2008,Divochiyetal2008,Gerritsetal2011,Calkinsetal2013,Chengetal2013,Mattiolietal2016,Guoetal2017,Zhuetal2020}, with certain architectures now capable of detecting up to 100 photons \cite{Chengetal2022arxiv,Eatonetal2022arxiv}, paving way for these quantum advantages.\footnote{Alternative schemes for inferring photon-number distributions without PNRDs are also available \cite{Zambraetal2005}.} What, then, occurs in the realistic situation that these detectors are imperfect? In particular, what happens when the detectors register spurious incident photons, which may come from another mode or intrinsic noise in the detector? This question has been addressed in the context of detectors capable only of discriminating between the presence and absence of photons \cite{Barbierietal2016}; we here study the important scenario of the effect of dark counts on PNRDs capable of discriminating between arbitrary numbers of photons.\footnote{PNRDs subject to dark counts have been studied in the context of estimating a relative phase \cite{Izumietal2016}}

There are a variety of PNRD architectures, each with different characteristics that motivate our study. Transition-edge sensors (TESs), for example, directly resolve the different amounts of energy imparted by different numbers of photons to infer a particular photon number. These devices must be extremely well calibrated and boast negligibly low dark count rates caused by intrinsic noise in the TES \cite{Milleretal2003}. However, the low dark count rates can be overwhelmed by any stray light impinging from an undesired mode, so these spurious noise counts must be accounted for in any investigation. Another example are multiplexed superconducting-nanowire detectors (SNDs), where incident light is split into a large number detectors capable of detecting the presence or absence of light and the photon number is inferred from the total number of detectors that fire \cite{Dauleretal2007}; moreover, an SND has recently been used to directly count up to four photons \cite{Zhuetal2020}. These then have to deal with dark count rates that are orders of magnitude larger than those of a TES, again motivating our study.
Photon-number-resolving cameras are also increasingly being used with a wide range of applications from characterizing unknown sources \cite{Rezaeeetal2022arxiv} to quantum imaging \cite{Wolleyetal2022}, with higher dark count rates than those of cryogenic detectors. Further use of these detection technologies in quantum sensing and metrology necessitates our study of dark counts in optimal transmission estimation.
We show how to analyze and maintain quantum advantages using realistic devices that are noisy due to anything other than the light they are trying to measure.

We study transmission estimation from the perspective of Fisher information, which quantifies the minimum uncertainty one can attain in estimating a transmission parameter. We use the Fisher information paradigm to compare classical (coherent) and quantum (Fock) probe states in their sensing performance using both ideal and realistic detectors to measure the transmitted light, through an analysis using the true photocounting statistics underlying noisy PNRDs. This allows us to demonstrate the marked quantum advantages of quantum probe light and how they vary with detector noise and other imperfections in realistic scenarios.

\subsection{Mathematical preliminaries}
Any transmission, reflection, or loss for a bosonic mode annihilated by $\ha$ can be mathematically described by the input-output relation
\eq{
    \ha\to\eta\ha+\sqrt{1-\eta^2}\hat{v}.
    \label{eq:loss transformation}
} This transmits some of the light to another mode, annihilated by $\hat{v}$, that is initially in its vacuum state and can account for the combined effect of  arbitrarily many transmissions, reflections, and losses. 
It even accounts for detector imperfections that are treated as loss, which cannot be distinguished from loss prior to the detector; for example, if the transmission before the detector is $\eta_0$ and the detector inefficiency is characterized by a transmission $\eta_{\mathrm{d}}$, the overall transmission only depends on the total effect $\eta=\eta_0\eta_{\mathrm{d}}$.
We refer to the real parameter $\eta$ as the transmission probability amplitude that we are trying to estimate, where the transmission probability is $\eta^2$. To find the overall effect on the mode of interest, we trace out the mode annihilated by $\hat{v}$, which tends to create a mixed state for most quantum input states \cite{Kimetal2002,Xiangbin2002,Asbothetal2005,Jiangetal2013,GoldbergJames2018,Fuetal2020,GoldbergHeshami2021}.

The crucial quantity for evaluating the power of a measurement procedure is the Fisher information. Given a probability distribution $\{p_m\}$ that depends on the parameter of interest, the Fisher information is defined as
\eq{
    \mathsf{F}(\eta;\{p_m\})=\sum_m p_m\left(\frac{\partial \ln p_m}{\partial \eta}\right)^2=\sum_m p_m^{-1}\left(\frac{\partial  p_m}{\partial \eta}\right)^2.
    \label{eq:Fisher information}
} It provides a lower bound to the ultimate precision with which any unbiased estimator can be constructed, quantified through the Cram\'er-Rao bound
\eq{
    \mathrm{Var}(\eta)\geq \mathsf{F}(\eta;\{p_m\})^{-1}.
} This decreases as $1/M$ when identical measurements are performed $M$ times, so we here consider the optimal single-shot ($M=1$) measurement. For a given probe, one can maximize the Fisher information over all possible measurement strategies to arrive at the quantum Fisher information, which has many useful properties \cite{SidhuKok2020} including methods for directly calculating without needing to manually optimize Eq. \eqref{eq:Fisher information}.

We define our input states in terms of a superposition over states with definite photon number (Fock states) $\ket{\psi}=\sum_{n\geq 0}\psi_n\ket{n}$, where $\ket{n}=\had^n\ket{\mathrm{vac}}/\sqrt{n!}$ and $\ha\ket{\mathrm{vac}}=0$.
Using the quantum Fisher information, one can show that the best setup for estimating the value of $\eta$ is to begin with a Fock state $\ket{N}$, apply the loss transformation of Eq. \eqref{eq:loss transformation}, then measure the photon-number distribution \cite{Adessoetal2009}
\eq{
p_m=\expct{\ket{m}\bra{m}}.
}
This achieves the maximum Fisher information of
\eq{
\mathsf{F}_{\ket{N}}(\eta)=4\frac{N}{1-\eta^2}.
\label{eq:Fisher opt Fock}
} In comparison, a classical input state with the same average energy $|\alpha|^2=N$,
$
    \ket{\alpha}=\eu^{-|\alpha|^2/2}\sum_{n=0}^\infty\frac{\alpha^n}{\sqrt{n!}}\ket{n},
$ only has a maximum (quantum) Fisher information of $4N$, so the quantum scheme outperforms the classical limit by the factor of $1-\eta^2$ using the same (optimal) measurement strategy. 

Now, when the detector has spurious counts such as dark counts, the measured photon-number distribution will differ from the state's underlying distribution. How does this affect the results? We find the true expected distribution in the presence of noise and detector imperfections. This allows us to quantify the Fisher information gleaned by a realistic detector, which we can again inspect for quantum advantages relative to classical input light. We demonstrate the quantum advantages attainable by Fock states with imperfect detectors in this important task of transmission sensing.

\section{Effects of dark counts}
\subsection{Photon-number distributions}
We know from photodetection theory that
\eq{
    \ket{m}\bra{m}=:\frac{\hat{n}^m\eu^{-\hat{n}}}{m!}:\,,
}where $\hat{n}=\had\ha$ and $:\cdot:$ is the normal-ordering operator that places all $\ha$s on the right side of all $\had$s. This facilitates the definition of an operator with spurious counts by replacing $\hat{n}$ with $\hat{n}+\nu$, where $\nu$ is the dark count rate \cite{Karpetal1970,Semenovetal2008,Sperlingetal2012}. The probability of registering $m$ photons is then given by
\eq{
    p_m=\expct{:\frac{(\hat{n}+\nu)^m\eu^{-\hat{n}-\nu}}{m!}:}=\eu^{-\nu}\sum_{k=0}^m \frac{\nu^{m-k}}{(m-k)!}\expct{\ket{k}\bra{k}}.
    \label{eq:photon number probability dark}
}

We can consider two probe states that have already been subject to loss:
\eq{
|\alpha\rangle\to \eu^{-|\eta\alpha|^2/2}\sum_{n=0}^\infty\frac{\eta^n\alpha^n}{\sqrt{n!}}\ket{n}=\ket{\eta\alpha}
}
 and
\eq{
    \ket{N}\bra{N}\to\sum_{n=0}^N\binom{N}{n}\eta^{2n} \left(1-\eta^2\right)^{N-n}\ket{n}\bra{n}.
} By calculating their true measured photon-number distributions using Eq. \eqref{eq:photon number probability dark}, we can evaluate the Fisher information from each distribution to see how well they truly perform in realistic conditions. 

Coherent states yield the true distribution
\eq{
    p_m(\ket{\alpha})&=\eu^{-\nu}\sum_{k=0}^m \frac{\nu^{m-k}}{(m-k)!}\eu^{-|\eta\alpha|^2}\frac{|\eta\alpha|^{2k}}{k!}\\
    &=\frac{(|\eta\alpha|^2+\nu)^m\eu^{-(|\eta\alpha|^2+\nu)}}{m!},
} which can also be directly inferred from the normal-ordered prescription. 
We can then use the definitions of the confluent hypergeometric function $U$ and the generalized Laguerre polynomials $L$ to find the true photon-number distributions for Fock states that have been subject to loss:
\eq{
    p_m(\ket{N})&=\eu^{-\nu}\sum_{k=0}^m \frac{\nu^{m-k}}{(m-k)!}\binom{N}{k}\eta^{2k} \left(1-\eta^2\right)^{N-k}\\
    &=\frac{\eu^{-\nu} (-\eta^2)^m \left(1-\eta^2\right)^{N-m} U\left(-m,-m+N+1,\frac{(\eta^2-1) \nu}{\eta^2}\right)}{m!}\\
    &=\eu^{-\nu} \eta^{2m} \left(1-\eta^2\right)^{N-m} L_m^{(N-m)}\left(\nu(1-\eta^{-2})\right),
    \label{eq:Fock state prob dist}
} where the binomial factor accounts for $m,k> N$ and there is always some nonzero probability of measuring any large photon number $m \gg N$.\footnote{Note similarities with the matrix elements of the displacement operator $D$ that enacts $D(\alpha)|\mathrm{vac}\rangle=\ket{\alpha}$: $\langle n|D(\xi)|m\rangle=\sqrt{\frac{m!}{n!}}\xi^{n-m}\eu^{-|\xi|^2/2}L_m^{(n-m)}(|\xi|^2)$ \cite{CahillGlauber1969}.}

An alternative method for obtaining the true underlying photon-number distribution may work better for arbitrary initial states. Instead of calculating Eq. \eqref{eq:photon number probability dark} after a probe has been subject to loss through Eq. \eqref{eq:loss transformation}, we can enact the substitution
$\hat{n}\to\eta^2\hat{n}+\nu$ and calculate expectation values in the states before loss has occurred.\footnote{A related procedure for adding spurious noise counts is to consider a bath of thermal photons instead of the vacuum operator in Eq. \eqref{eq:loss transformation}; this has recently been studied in the context of optimal transmission sensing in Refs. \cite{Gongetal2021,Gongetal2022arxiv} and converges to the present procedure $\hat{n}\to\eta^2\hat{n}+\nu$ in the limit of a large number of thermal modes \cite{Semenovetal2008}.}
For a general state that has not yet been subject to loss, we thus find the probability distribution including loss and dark counts to be
\eq{
    p_m&=\expct{:\frac{(\eta^2\hat{n}+\nu)^m\eu^{-\eta^2\hat{n}-\nu}}{m!}:}\\
    &=\frac{\eu^{-\nu}}{m!}\sum_{k=0}^m\binom{m}{k}\eta^{2k}\nu^{m-k}\sum_{l=0}^\infty \frac{(-\eta^2)^l}{l!}\expct{:\had^{k+l}\ha^{k+l}:}.
} Pure states then yield a convex combination of the Fock-state results from Eq. \eqref{eq:Fock state prob dist}:
\eq{
    p_m(\ket{\psi})&=\frac{\eu^{-\nu}}{m!}\sum_{k=0}^m\binom{m}{k}\eta^{2k}\nu^{m-k}\sum_{l=0}^\infty \frac{(-\eta^2)^l}{l!}\\
    &\qquad\times\sum_n|\psi_n|^2\frac{n!}{(n-k-l)!}\\
    &=\frac{\eu^{-\nu}}{m!}\sum_n|\psi_n|^2\sum_{k=0}^m\binom{m}{k}\eta^{2k}\nu^{m-k}\frac{(1-\eta^2)^{n-k}}{(n-k)!}n!\\
    &=\eu^{-\nu}\sum_n|\psi_n|^2\left(1-\eta^2\right)^{n-m}  \left(\eta^2\right)^{m} L_m^{(n-m)}\left(\nu-\frac{\nu}{\eta^2}\right).
} The results for mixed-state inputs are similarly given by convex combinations of the results for pure-state inputs.

\subsection{Fisher information}
When the spurious count rate $\nu$ is known \textit{a priori}, an estimate of $\eta$ provides the Fisher information as in Eq. \eqref{eq:Fisher information}. If, however, $\nu$ is not known, one must treat it as a nuisance parameter \cite{Basu1977,Suzuki2020,Suzukietal2020}. This is done by extending the Fisher information into a symmetric matrix with components
\eq{
    \mathsf{F}_{ij}=\sum_m\frac{1}{p_m}\frac{\partial p_m}{\partial\theta_i}\frac{\partial p_m}{\partial\theta_j}
} and using it to provide a lower bound on the covariance matrix
\eq{
    \mathrm{Cov}(\theta_i,\theta_j)\geq\left(\mathsf{\mathbf{F}}^{-1}\right)_{ij}.
}
This serves to only increase the minimum uncertainty, with
\eq{
    \mathrm{Var}(\eta)\geq \frac{1}{\mathsf{F}_{\eta\eta}-\mathsf{F}_{\eta\nu}^2/\mathsf{F}_{\nu\nu}}.
}

\subsubsection{Coherent-state (classical) inputs}
We start by considering $\nu$ to be known. For coherent states, we have
\eq{
    \partial_\eta p_m(\ket{\alpha})=2\eta|\alpha|^2p_m(\ket{\alpha})\left(\frac{m}{\eta^2|\alpha|^2+\nu}-1\right),
} yielding
\eq{
    \mathsf{F}_{\eta\eta}(\ket{\alpha})=\frac{4 |\alpha|^4 \eta^2}{|\alpha|^2 \eta^2+\nu}.
} We again write the average input energy as $|\alpha|^2=N$ to show how the minimum uncertainty decreases \eq{
    \mathrm{min}[\mathrm{Var}(\eta)]=\frac{1}{4N}\to\frac{1}{4N}+\frac{\nu}{\eta^2N^2}.
} Intuitively, the extra uncertainty contributed by dark counts increases for increased dark count rates $\nu$ and decreases for increased transmission probability $\eta^2$ and increase probe state energy $N$; small transmissions and weak coherent states are overwhelmed by dark counts.

Could we use this setup to instead \textit{measure} the dark counts with coherent states? If the loss parameter were known \textit{a priori}, we would use
\eq{
    \mathsf{F}_{\nu\nu}(\ket{\alpha})=\frac{1}{\eta^2N+\nu}.
} These dark counts would then be easier to estimate by using a weak coherent state or a setup with a small transmission probability relative to the dark count rates ($\eta^2N\ll \nu$). 

The final element required for evaluating this estimation procedure using the tools of nuisance parameters is the off-diagonal term
\eq{
\mathsf{F}_{\eta\nu}=\mathsf{F}_{\nu\eta}=\frac{2 |\alpha|^2 \eta}{|\alpha|^2 \eta^2+\nu}.
} This makes the Fisher information matrix $\mathsf{\mathbf{F}}$ singular -- one measurement cannot be used to simultaneously measure both of these two parameters! It is \textit{impossible} to measure the loss parameter using coherent state inputs when nothing is known \textit{a priori} about the dark count rate and vice versa. This is, in fact, similar to the problem of doing the measurement when the input coherent state strength is not known, because the output cannot distinguish between loss and a weaker input. One can see this directly by noting that the probability distribution only depends on $(|\alpha|, \eta, \nu)$ through a single functional form $\eta^2|\alpha|^2+\nu$, so it is of course impossible to tease the parameters apart with this measurement strategy alone, as is known from studies of singular Fisher information matrices \cite{StoicaMarzetta2001,XavierBarraso2004,Goldbergetal2021singular}.

How would this change if we had some \textit{a priori} information? Say that we already gained some Fisher information $\mathsf{f}$ about $\nu$ in an alternate experiment. Then, since Fisher information is additive, we would find 
\eq{
    \mathsf{\mathbf{F}}(\ket{\alpha};\eta,\nu)=\frac{1}{\eta^2N+\nu}\begin{pmatrix}
    (2N\eta)^2&2N\eta\\
    2N\eta&1
    \end{pmatrix}+\begin{pmatrix}0&0\\0&\mathsf{f}
    \end{pmatrix}.
} Inspecting the $\eta\eta$ element of the inverse, which gives us the minimum uncertainty for estimating $\eta$, we find
\eq{
    \mathrm{Var}(\eta)\geq& \frac{1}{\mathsf{F}_{\eta\eta}-\mathsf{F}_{\eta\nu}^2/(\mathsf{F}_{\nu\nu}+\mathsf{f})}=\frac{1}{4N}+\frac{\nu}{4\eta^2N^2}+\frac{1}{4N^2\eta^2 \mathsf{f}}.
} The extra \textit{a priori} information $\mathsf{f}$ about $\nu$ diminishes the uncertainty on $\eta$ to its theoretical minimum in the presence of dark counts, and helps get there fast with stronger coherent states and larger transmission. With tiny transmission, more \textit{a priori} information about dark counts is required to well estimate the transmission parameter using coherent probe states.

Another idea would be to repeat this experiment twice with two different coherent state energies $qN$ and $(1-q)N$. The total Fisher information would be
\eq{
    \mathsf{\mathbf{F}}(\ket{\alpha};\eta,\nu)=&\frac{1}{\eta^2qN+\nu}\begin{pmatrix}
    (2qN\eta)^2&2qN\eta\\
    2qN\eta&1
    \end{pmatrix}\\
    &+\frac{1}{\eta^2(1-q)N+\nu}\begin{pmatrix}
    (2(1-q)N\eta)^2&2|\alpha|^2\eta\\
    2(1-q)N\eta&1
    \end{pmatrix},
} yielding the minimum uncertainty
\eq{
    \mathrm{Var}(\eta)\geq& \frac{\eta^2 N+2 \nu}{4 \eta^2 N^2 (1-2 q)^2}\geq \frac{1}{4N}+\frac{\nu}{2\eta^2N^2}.
} The minimum is obtained when $q=0$ or $q=1$, which makes sense given that the value of $\nu$ is best estimated when $N=0$ and that of $\eta$ when $N$ is maximal, and this minimum is equivalent to the case where the \textit{a priori} information known about $\nu$ is $\mathsf{f}=1/\nu$.
In fact, if the figure of merit is the amount of energy used in the probe, one can get an unlimited amount of information about the dark count rate $\nu$ by collecting Fisher information $1/\nu$ for zero input probe energy and repeating the process a large number of times. This helps motivate the conclusion that, in practice, $\nu$ tends to be known \textit{a priori}.

\subsubsection{Fock states}
The first difference we notice between coherent-state inputs and Fock-state inputs is that $p_m(\ket{N})$ depends differently on $\eta$, $N$, and $\nu$, so there is no fundamental barrier to simultaneously estimating any of these in a single measurement with a Fock state. Still, to compare with coherent states, we might as well assume the parameters other than $\eta$ to be known \textit{a priori}, only focusing on the $\mathsf{F}_{\eta\eta}$ element of the Fisher information matrix.

The Fisher information is given by the sum over all $m$ of the positive terms
\begin{widetext}
\eq{
    \frac{1}{p_m}\left(\frac{\partial p_m}{\partial \eta}\right)^2=\frac{4 \eu^{-\nu} \eta^{2 m-6} \left(1-\eta^2\right)^{-m+N-2} \left[\eta^2 \left(m-N \eta^2\right) L_m^{(N-m)}\left(\nu-\frac{\nu}{\eta^2}\right)+\nu \left(\eta^2-1\right) L_{m-1}^{(N-m+1)}\left(\nu-\frac{\nu}{\eta^2}\right)\right]{}^2}{L_m^{(N-m)}\left(\nu-\frac{\nu}{\eta^2}\right)}.
    \label{eq:Fisher terms Fock}
}
\end{widetext}
We can analytically compute the sum for the two terms in the expanded square that do not have a Laguerre polynomial in the denominator, leaving us with
\eq{
    \mathsf{F}_{\eta\eta}&(\ket{N})=\frac{4 \left[ \left(\eta ^2-\eta ^4\right) N-\nu  (\nu +1)\right]}{\eta ^2 \left(1-\eta ^2\right)^2}\\
    &+4 e^{-\nu} \nu^2\sum_{m=0}^\infty \eta^{2 m-6} \left(1-\eta^2\right)^{N-m} \frac{  L_{m-1}^{(N-m+1)}\left(\nu-\frac{\nu}{\eta^2}\right){}^2}{L_m^{(N-m)}\left(\nu-\frac{\nu}{\eta^2}\right)}.
} 
In the small-$\nu$ limit, we see the beginning of the deviation from  Fock states being ideal as before:
\eq{
    \mathsf{F}_{\eta\eta}(\ket{N})=\frac{4N}{1-\eta^2}-\frac{4}{\eta ^2 \left(1-\eta ^2\right)^2}\nu+\mathcal{O}(\nu^2),
} while the large-$\nu$ limit (which is normally avoided in reasonable experiments) shows a behaviour independent from $N$ yet retains the enhanced scaling with large transmissibility:
\eq{
    \mathsf{F}_{\eta\eta}(\ket{N})=\frac{4 \left[ \left(\eta ^2-\eta ^4\right) N-\nu  (\nu +1)\right]}{\eta ^2 \left(1-\eta ^2\right)^2}+\mathcal{O}\left(\frac{1}{\nu}\right).
}

To continue, we must analyze the Fisher information for Fock states numerically.
Since each term $\frac{1}{p_m}\left(\frac{\partial p_m}{\partial \eta}\right)^2$ in Eq. \eqref{eq:Fisher terms Fock} is positive, truncating the sum at some $m$ will always provide less information than truncating it at a larger $m$. This directly implies that PNRDs capable of resolving more photons will always provide more information than those capable of resolving fewer photons. We perform this truncation at a sufficiently large number such that we do not have to worry about edge effects where more photons arrive at the detector than can be distinguished by such; this large number is determined by the total intensity $N+\nu$ that a detector might register. Fock states up until $N\approx 4-8$ have been experimentally generated \cite{Tiedauetal2019,Engelkemeieretal2021arxiv,Thekkadathetal2020} and dark count rates tend to be much smaller than unity, so current PNRD technologies capable of resolving 10-100 photons should all suffice to avoid edge cases. In the remaining work we truncate our sums at $m=30$.

\begin{figure*}
    \centering
    \includegraphics[trim={0.8cm 0 0 0},clip]{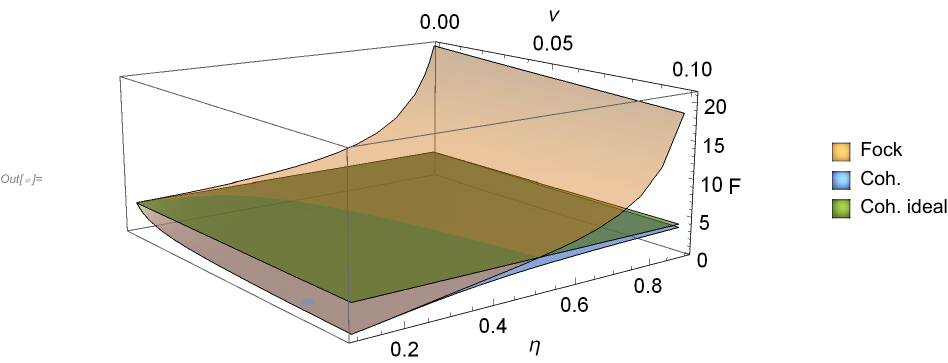}
    \caption{Fisher information for estimating a transmission parameter $\eta$ with PNRDs capable or resolving up to 30 photons versus dark count rate $\nu$ and transmission probability amplitude $\eta$. All of the input states have average energy $N=1$. Fock states always outperform coherent states, with Fisher information increasing as $1/(1-\eta^2)$ for large $\eta$ to present a quantum advantage. We also include for comparison the performance of coherent states not subject to spurious dark counts, the planar curve in the image, to show that Fock states subject to spurious counts still outperform ideal coherent states. Coherent states with dark counts interpolate between the same scaling with $\eta$ as Fock states for small $\eta$ and the same behaviour as coherent states with no dark counts for large $\eta$.}
    \label{fig:Fock vs coh n1}
\end{figure*}

We first compare Fock states' performance to that of coherent states for different values of the transmission parameter $\eta$ and the dark count rate $\nu$. In Fig. \ref{fig:Fock vs coh n1}, we can see how Fock states with $N=1$ always outperform coherent states with $|\alpha|^2=1$ in the presence of dark counts for all values of $\nu$ and $\eta$, extending their advantages from the $\nu=0$ case. We can also see that Fock states even outperform coherent states when the coherent states are not exposed to dark counts, so long as the dark count rates are sufficiently low and the transmission probability is sufficiently high; such a comparison is required when one compares the use of a single strong coherent state against a multitude of single-photon states with total comparable energy. This behaviour is reproduced for all $N$ that we evaluated. 

\begin{figure*}
    \centering
    \includegraphics[trim={0.8cm 0 0 0},clip]{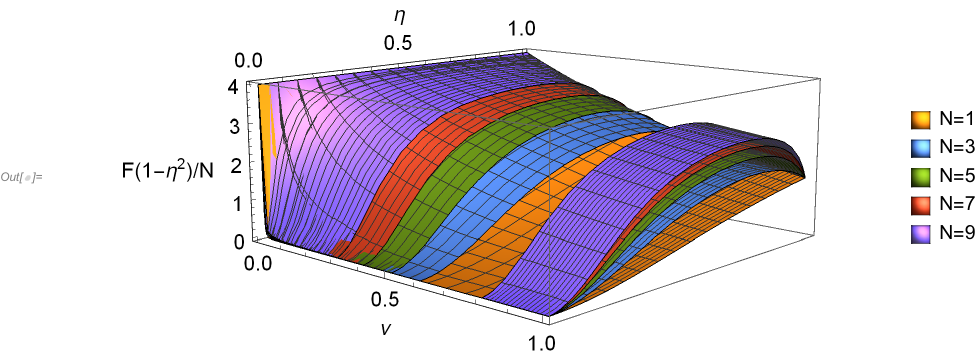}
    \caption{Scaled Fisher information for estimating a transmission parameter $\eta$ with PNRDs capable or resolving up to 30 photons versus dark count rate $\nu$ and transmission probability amplitude $\eta$. We have cut out sections of the curves (from $\nu=0.6-0.8$ for $N=5$, $\nu=0.5-0.8$ for $N=3$, $\nu=0.4-0.8$ for $N=7$, and $\nu=0.3-0.8$ for $N=9$) to see their nested structures. All of the input states are Fock states with various energies $N$ and have Fisher information favourably growing as $1/(1-\eta^2)$, so we scale the Fisher information by $(1-\eta^2)/N$ to inspect the advantages per photon. Fock states with more photons perform the best in the realistic scenario where dark counts are nonzero.}
    \label{fig:comparing Fock}
\end{figure*}

All Fock states retain their superior scaling with $1-\eta^2$ relative to coherent states. In contrast to the case with zero dark counts, however, different Fock states present different advantages. To compare the various Fock states, we scale their Fisher informations by $1-\eta^2$ and by $N$ to directly inspect their relative performance in Fig. \ref{fig:comparing Fock}. It is then clear that, in the presence of nonzero dark counts, larger Fock states with more energy provide the most information for a given amount of energy, breaking the equivalence between different Fock states in the $\nu=0$ regime of Eq. \eqref{eq:Fisher opt Fock} and helping motivate the generation of larger Fock states \cite{Thekkadathetal2020}.

\section{Conclusions}
We have performed a detailed comparison between coherent (classical) and Fock (quantum-optimal, definite-photon-number) states for sensing a transmission parameter $\eta$ using realistic photon-number-resolving detectors that are subject to spurious counts such as dark counts. The dramatic advantages of Fock states over coherent states is retained in this realistic scenario, where now Fock states with more energy are superior to weaker Fock states. This result helps spur the production of Fock states with more photons and of detectors capable of resolving these large numbers of photons, with applications in the quantum-enhanced sensing of a host of different effects.

\begin{acknowledgments}
    The authors acknowledge that the NRC headquarters is located on the traditional unceded territory of the Algonquin Anishinaabe and Mohawk people. The authors also acknowledge support from NRC's Quantum Sensing Challenge program. AZG acknowledges funding from the NSERC PDF program.
\end{acknowledgments}

\end{document}